**Title: Giant Asymmetric Radiation from an Ultrathin Bianisotropic Metamaterial**

*Liang Peng\*, Kewen Wang, Yihao Yang, Yuntian Chen, Gaofeng Wang, Baile Zhang\*, and Hongsheng Chen\**

Prof. L. Peng, K. Wang, Prof. G. Wang
Key Laboratory for RF Circuits and Systems, Hangzhou Dianzi University,
Ministry of Education, Hangzhou, 310018, China
E-mail: pengl@hdu.edu.cn

Dr. Y. Yang, Prof. H. Chen
State Key Laboratory of Modern Optical Instrumentation,
Zhejiang University, Hangzhou, 310027, China
E-mail: hansomchen@zju.edu.cn

Prof. Y. Chen
School of Optical and Electronic Information,
Huazhong University of Science and Technology, Wuhan, 430074, China

Prof. B. Zhang
Division of Physics and Applied Physics, School of Physical and Mathematical Sciences,
Nanyang Technological University, Singapore 637371, Singapore
E-mail: blzhang@ntu.edu.sg

Unidirectional radiation is of particular interest in high-power lasing and optics. Commonly, however, it is difficult to achieve a unidirectional profile in such a system without breaking reciprocity. Recently, assisted by metamaterials without structural symmetry, antennas that radiate asymmetrically have been developed, hence providing the possibility of achieving unidirectionality. Nevertheless, it has been challenging to achieve extremely high radiation asymmetry in such antennas. Here, we demonstrate that this radiation asymmetry is further enhanced when magnetic plasmons are present in the metamaterials. Experimentally, we show that a thin metamaterial with a thickness of approximately $\lambda_0/8$ can exhibit a forward-to-backward emission asymmetry of up to 1:32 without any optimization. Our work paves the way for manipulating asymmetric radiation by means of metamaterials and may have a variety of promising applications, such as directional optical and quantum emitters, lasers, and absorbers.

## 1. Introduction

In electromagnetics and optics, the well-known metasurfaces could have the ability of manipulating the profiles of electromagnetic (EM) waves, including the control of wave front, polarizations, holograms, angular momentums, etc.[1-8] Commonly, metamaterials (MMs) with asymmetric structures are known to exhibit asymmetric (or bianisotropic) EM responses, such as single-sided radiation and absorption patterns and unbalanced forward-backward emission.[9-26] In particular, asymmetric forward-backward radiation is of great interest in many photonic applications, such as high-power lasers and light detection.[27-34] To enhance this asymmetry, MM structures are often optimized through numerical approaches. Recently, the existence of fundamental bounds on the overall responses of these structures has been reported based on temporal coupled-mode theory[35] and the principle of time-reversal symmetry (TRS).[36] Nevertheless, under both schemes, it has been shown that unbounded asymmetric responses may occur when the structure is totally reflecting. However, in all of the literature cited above, total reflection does not guarantee the occurrence of extreme asymmetry. The physics behind these exotic phenomena and the key to achieving maximal radiation asymmetry still remain unrevealed.

For MMs consisting of asymmetric constituents, magneto-electric (ME) resonances arise in the band of interest. Macroscopically, the ME effect in homogeneous MMs can be described in terms of bianisotropy.[37,38] For a MM thin film, or a metasurface, a ME tensor may be defined.[26] Along with classical electromagnetics, these two approaches are equally effective in describing the collective EM responses of MMs. Here, without loss of generality, we first adopt an effective medium description and then extend the main results to ultra-thin MMs. We show that plane waves propagating in opposite directions may have dramatically different field profiles in the interiors of bianisotropic MMs, due to which systematic asymmetry is induced in the MMs' EM responses. Analytically, with enhancing its macroscopic bianisotropy, the MM becomes totally reflective, which provides the possibility of unbounded system asymmetry. With the aid of a bianisotropic bulk MM, the radiation of a short dipole can be controlled. It is shown that although asymmetric radiation is indeed induced due to the bianisotropy of the bulk material, the maximal asymmetry may be achievable only when the MM's permeability tends to zero, i.e., in the presence of magnetic plasmons. The equivalence between bianisotropy and ME polarization holds for both bulk MMs and thin-film structures. Computer-based simulations and experimental measurements verify our findings.

## 2. Asymmetric Behaviour of Bianisotropic MMs

We start by studying the EM responses of standard split-ring resonators (SRRs). SRRs are often applied in MMs to excite strong magnetic resonance in order to achieve negative permeability.[39,40] However, if the SRRs are not geometrically symmetrical, an ME effect will be induced, and the effective medium can be described in terms of macroscopic bianisotropy,[41] as in the case of the unbalanced SRRs (USRRs) shown in **Figure** 1(a).

In essence, MMs with asymmetric constituents exhibit asymmetrical EM responses to waves coming from opposite directions. Figure 1(b) shows a passive antenna system composed of a short dipole combined with a bianisotropic MM reflector, which can radiate asymmetrically. In particular, in the case that the MM is totally reflective, single-sided radiation may be realizable. There are basically two combinations of the dipole and the MM reflector, depending on the spatial orientation of the SRRs, which are illustrated by Case A and B in Figure 1(b). Because the bounds for the slab's asymmetric EM response is broken by the total reflection,[36] the radiation capability of the antennas in Case A and B would be rather different (as the so-called asymmetric radiation in this paper), as we show later. Here, before we proceed with a detailed discussion of radiation asymmetry, we study the EM mechanism for an individual USRR to gain a better understanding of the macroscopic phenomenon. In Figure 1(c), a USRR is placed in free space with plane wave illuminating. Basically, by breaking the structural symmetry, the capacitive part and the inductive part of the USRR are physically separated. Under resonance conditions, the localized electric and magnetic fields inside a USRR is strongly enhanced and non-uniform, with little overlapping. As a result the network effect between the capacitive and the inductive parts is not negligible, regardless of the physical size of the resonator. For instance, in Figure 1(c), the USRR shows stronger capacitive behaviour for the wave coming from the left side than for the one coming from the right because the capacitive opening can more easily interact with the former than the latter. For better interpretation, a simple transmission line model is used to explicate the physical picture of this asymmetrical effect. Although the purely passive circuit model is not able to describe the USRR's ME effect,[9] we learn that the network effect makes the USRR behave dramatically different in responding different external excitations.

To explore the macroscopic EM responses, we turn to the source-free Maxwell equations. For the MM shown in Figure 1(a), the constitutive relations are written as $\overline{D} = \overline{\overline{\varepsilon}} \cdot \overline{E} + \overline{\overline{\chi}} \cdot \overline{H}$ and $\overline{B} = \overline{\overline{\mu}} \cdot \overline{H} - \overline{\overline{\chi}}^T \cdot \overline{E}$, where $\overline{\overline{\chi}}$ is the ME tensor. For lossless and reciprocal media, $\overline{\overline{\chi}}$ is purely imaginary.[38] In Figure 1(a), it is seen that only $H_x$ and $E_z$ are coupled by the USRRs; hence, the effective parameters can be expressed as $\overline{\overline{\varepsilon}} = diag[\varepsilon_x, \varepsilon_y, \varepsilon_z]$, $\overline{\overline{\mu}} = diag[\mu_x, \mu_y, \mu_z]$, and $\overline{\overline{\chi}} = i \begin{bmatrix} 0 & 0 & 0 \\ 0 & 0 & 0 \\ \chi & 0 & 0 \end{bmatrix}$. We note that for the dispersive MM to be treated as an effective medium, the band of interest should be far beyond the MM's resonance so that the effective parameters may be reasonably defined. Classically, the response of a homogeneous MM can be well described by means of plane waves. For the MM depicted in Figure 1(a), we assume that a *z*-polarized plane wave is propagating along the *y* direction. This assumption is adopted so that $\bar{H}$ is *x*-polarized; then, the bianisotropy takes effect for the forward ($\hat{k} = +\hat{y}$) and backward ($\hat{k} = -\hat{y}$) waves. Let $\bar{k} = -\hat{y}k$, $\bar{E} = \hat{z}E_z$, and $\bar{H} = \hat{x}H_x$; by solving the source-free Maxwell equations, we obtain $k^2 = \omega^2(\varepsilon_z \mu_x - \chi^2)$. Meanwhile, the intrinsic wave impedance can be well defined as $\eta = \frac{E_z}{H_z}$; in greater detail, this is evaluated as $\eta = \frac{\omega \mu_x}{k + i\omega\chi}$, or $\eta = \frac{k - i\omega\chi}{\omega \varepsilon_x}$.

We see that if $\chi = 0$, causing the bianisotropy to disappear, then the wave impedance $\eta$ returns to the conventional case. However, if $\chi \neq 0$, then $\eta$ will differ depending on the direction of propagation; i.e., $\eta_\pm = \frac{\omega \mu_x}{\pm k + i\omega\chi}$ or $\eta_\pm = \frac{\pm k - i\omega\chi}{\omega \varepsilon_x}$, where $\eta_+$ and $\eta_-$ are the impedances for $\hat{k} = \pm \hat{y}$, respectively. Two cases should be addressed: (A) For propagating waves, *k* is real, and $\eta_+ = -\eta_-^*$. As a result, waves travelling in the $\pm \hat{y}$ directions behave symmetrically,[10] and consequently, the system's asymmetry is bounded.[36] (B) For evanescent waves, *k* is imaginary or complex (for lossy materials), and $|\eta_+| \neq |\eta_-|$. This indicates that waves travelling in the $\pm \hat{y}$ directions are no longer symmetrical, despite their identical decay rates. Although asymmetry arises, such passive MMs have no reason to violate the reciprocity principle. The sole option for achieving both asymmetry and reciprocity is for waves coming from opposite directions to interact with the MM in an isolated manner.[36] This isolation causes the MM to totally reflect, which is a sign of bound breaking.[35,36] Interestingly, this isolation begins at *k*=0, i.e., where $\eta_+$ and $\eta_-$ are purely imaginary and equal, in which case the polarization/magnetization of the MM cannot distinguish among waves coming from different directions.

**3. Asymmetric Radiation Assisted by a Bulk MM**

To realize asymmetric radiation, a current source is placed in front of a bulk MM, as in Cases A and B depicted in Figure 1(b). Because the bulk MM is totally reflective in both cases, the radiation from the current source ($J$) can be evaluated by integrating the Green's function in the half space, which is a rigorous procedure but rather complicated. Since the radiation in both cases is dominated by waves propagating in directions nearly normal to the interface, it is not necessary to fully solve the radiation problem to evaluate the radiation asymmetry. Here, as an alternative approach, we estimate the radiation asymmetry by considering only the normal-incidence interactions for simplicity. We note that although this approach is not rigorous, it is effective and reliable in approximating the radiation asymmetry, as shown below.

The electric field radiated in the normal direction can be expressed as $\overline{E}_t = \overline{E}_i + \overline{E}_r$, where $\overline{E}_i$ is the field excited directly from the current source and $\overline{E}_r$ is the field reflected by the bulk MM. From classical electromagnetics,[38,42] it can be conveniently found that $\overline{E}_r = \frac{\eta - \eta_0}{\eta + \eta_0} \overline{E}_i$, where $\eta$ is the intrinsic wave impedance inside the MM and $\eta_0$ is that in free space. By evaluating both Cases A and B, the asymmetry of the radiated power can be roughly estimated as follows:

$$A_R \approx \frac{|E_{t,B}|^2}{|E_{t,A}|^2} = \left| \frac{\mu_x + i\eta_0(\sqrt{|\varepsilon_z \mu_x - \chi^2|} - \chi)}{\mu_x + i\eta_0(\sqrt{|\varepsilon_z \mu_x - \chi^2|} + \chi)} \right|^2, \quad (1)$$

where $i = -\sqrt{-1}$.

As seen from Equation (1), the radiation asymmetry is dominated by $\chi$; i.e., a non-zero $\chi$ appears to be essential for inducing asymmetry, and extreme asymmetry ($A_R \to 0$) may be achieved by means of infinite bianisotropy ($\chi \to \infty$), provided that all other parameters are finite. However, this requirement is essentially unachievable, as $\chi$ must be finite in a practical system due to causality and the inevitable occurrence of loss.[43] An alternative way to achieve extreme asymmetry is to have $\mu_x = 0$ in Equation (1), in which case $A_R$ may again approach a minimum ($\approx 0$). Hence, in practical scenarios, magnetic plasmons, rather than bianisotropy, will dominate the asymmetry. It is also worth noting that if $\chi$ is positive, then $A_R$ is always a number smaller than unity, and this bound is broken if the sign of $\chi$ is reversed, which means that the MM's (or USRRs') spatial orientation is flipped. In the remaining part of the paper, we present numerical simulations and experimentally measured results.

## 4. MM Design and Full-Wave Simulations

In this section, we present the design of the MM and the derivation of the relevant effective parameters. The basic USRR structure is shown in **Figure** 2(a). In this design, the complex structure is made of copper with a conductivity of $5.8 \times 10^7$ S/m. To facilitate fabrication in practice, the USRR is designed on a 1 mm thick FR4 substrate, which has a dielectric constant of 4.3 and a loss tangent of 0.025. For the remaining parameters, please refer to Figure 2(a).

To extract the MM's effective parameters, transmittance and reflectance (i.e., the S-parameters) from a slab consisting of one layer of USRRs are obtained through full-wave simulations, which are shown in Figure 2(b). The MM's resonance occurs in a range from below 4 GHz to above 5 GHz. At frequencies above 4 GHz, $|S_{11}|$ and $|S_{22}|$ differ, which is a sign of asymmetrical EM responses for lossy materials; hence, enhanced bianisotropy is expected. By applying the retrieval approach,[44,45] the effective parameters can be reasonably derived; the results are presented in Figure 2(c). In Figure 2(d), the effective index of refraction in the *y* direction is shown. In the band of interest, the wave number is almost purely imaginary; hence, total reflection occurs. In addition, high contrast in the intrinsic impedance is seen in the band from 4.2 GHz to 5 GHz; i.e., this is a band with enhanced bianisotropy. The effective wave impedances ($\eta_+$ and $\eta_-$) evaluated from the retrieved parameters are shown in Figure 2(e). As the MM becomes totally reflective, the intrinsic impedances for waves travelling in the $\pm \hat{y}$ directions begin to differ dramatically. Thus, the asymmetric EM propagation inside the MM bulk is well captured by $|\eta_+| \neq |\eta_-|$.

In the following, we present full-wave simulations of the asymmetric radiation generated by applying the proposed bianisotropic MM. To form a bulk MM of finite size, the unit cell periods used in the *x*, *y* and *z* directions are 16, 8 and 15, respectively. Schematic illustrations of the bulk MM are shown in **Figure** 3(a) and (b). In Figure 3(a), a short current (*J*) is applied on the left side of the bulk MM, where the openings in the USRRs appear (Case A). In Figure 3(b), *J* is applied on the right side; this scenario is labelled as Case B. According to our previous analysis, the MM shows high impedance in Case A, so enhanced radiation may be expected. This is very similar to the case in which a dipole is placed in the vicinity of a magnetic conductor of infinite impedance. By contrast, the radiation in Case B is suppressed by the low-impedance backsides of the USRRs. As a result, asymmetric radiation is induced.

In the simulations, a realistic short dipole made of copper was considered, with a length of $l_D$ =20 mm, and a short section of transmission line (with a length of $l_M$ =10 mm) was applied between the dipole and the 50 Ω feed port to make the dipole system resonate in the range between 4 GHz and 5 GHz. The structure of the dipole system is shown in the inset between Figure 3(a) and (b). Considering the dipole's short length, the current distribution will be simple and will be similar regardless of the backside load.[42] As a result, the influence of the bulk MM will be directly reflected in the radiation resistance, or, equivalently, the antenna's return loss. In Figure 3(c), we present the return losses in three cases, i.e., the dipole alone, Case A and Case B. We see that the dipole system is resonant in the range from 4 GHz to 5 GHz, but it is not well matched to the feed. In Case A, the return loss is greatly improved near 4.8 GHz, indicating that the radiation is enhanced due to the high-impedance backside of the MM. In Case B, the return loss from the antenna system becomes worse because of the destructive influence of the low-impedance backside of the MM. We further simulated the radiation patterns in Cases A and B at 4 GHz and 5 GHz, where extremely low and high impedances are involved, respectively; see Figure 3(d) and (e). At both frequencies, the radiation in Case A is more efficient than that in Case B. Moreover, the contrast between the radiated power patterns is higher at 4 GHz than at 5 GHz. To facilitate interpretation, the power gain contrast between Cases B and A is plotted in Figure 3(f). The simulated results are inconsistent with the theoretical curve in a range around 4.6 GHz, indicating that our theory does not yield precise estimates at those frequencies. However, for both the simulated and calculated curves, a dip in the radiation asymmetry is observed near 4 GHz, where magnetic plasmons occur ($\mu_x \approx 0$), and the two curves coincide throughout most of the frequency range.

## 5. Experiments for Antennas Assisted by a Bulk MM

Experimental radiation measurements were performed in a microwave anechoic chamber, as shown in **Figure** 4(a). The fabricated dipole and the bulk MM are depicted in the insets of Figure 4(a). The length of the transmission line section ($l_M$) was tuned so that the oscillation of the dipole system would fall into the band of interest, approximately near 4.5 GHz. First, the transmittance of the bulk MM was recorded; the results are shown in Figure 4(b). In the frequency range from 3.8 GHz to 5.4 GHz, the transmission is significantly degraded by the presence of the bulk MM; this forbidden band is roughly consistent with the simulation-based results presented in Figure 2.

The return losses recorded for the various system configurations are shown in Figure 4(c). In the absence of any back reflectors, the dipole system is mismatched. In Case A, the matching is significantly improved, and most of the power is radiated, whereas in Case B, the matching is degraded. These observed phenomena are fully consistent with our simulations. The radiation patterns in Cases A and B were also measured, with the results shown in Figure 4(d) and (e). At 4 GHz, the gain contrast measured between Cases A and B exceeds 10 dB, whereas this contrast is weaker at 5 GHz, as shown in Figure 4(e). In both Figure 4(d) and (e), it is obvious that the MM reflector assists in generating asymmetric radiation, and the asymmetry holds throughout the entire band of interest. The radiation asymmetry extracted from the measured results is plotted in Figure 4(f). The measured curve is roughly consistent with the simulated results and with the results calculated using Equation (1). For all curves, the maximal radiation asymmetry is observed near 4 GHz, and the radiation asymmetry remains below unity; both findings validate our theory.

## 6. Extension to Thin MM Structures

For the previous analysis, the radiation asymmetry was defined for the case in which a bulk MM is used as a back reflector. Since the physics behind this phenomenon is that each single unit cell structure responds to waves from opposite directions in different manners, we expect that this effect should also hold in other optical systems, such as metasurfaces,[46-48] gratings and thin films consisting of only one layer of asymmetric constituents.

To confirm this expectation, additional simulations and experiments were conducted. In **Figure** 5(a), the design of an antenna consisting of only one layer of USRRs is presented. The thickness of the MM is 8 mm, which is roughly equal to $\lambda_0/8$ (where $\lambda_0$ is the free-space wavelength) in the band of interest. A short current (*J*) is placed at the centre of the layer. Considering the layer's thickness and its periodicity in the *x* and *z* directions, there is no obvious boundary for waves coming from the $\pm \hat{y}$ directions; as a result, the dipole's radiation will be dominated by the ME polarizability of the USRRs. In Figure 5(b), the fabricated antenna and its experimental setup are shown. For the measurements, the system was fed by a monopole connected to the coaxial feed cable and was kept stable with the aid of a foam slab. In Figure 5(c), the simulated and experimentally measured radiation asymmetry profiles are shown. Extreme asymmetry is observed near 4 GHz in the simulation, but this phenomenon shifts to approximately 4.2 GHz in the experimental data. This shift in the operating frequencies may be due to fabrication imperfections, such as the use of a non-ideal substrate. According to both the simulated and measured results, the radiation asymmetry dips to below 0.031 (equivalent to 1:32), indicating nearly unidirectional radiation, as expected.

## 7. Conclusion

In conclusion, we have demonstrated the asymmetric responses of MMs without geometrical symmetry. The physical mechanism of this phenomenon is the spatial separation of the capacitive and inductive components, as well as the network effect in oscillating structures. The behaviour of the entire system is significantly influenced, or even determined, by which of these components dominates the response to external illumination. With enhanced ME resonance, bianisotropic MMs would become totally reflective and respond differently to incoming waves from opposite directions, which provide the possibility of asymmetric radiation control. With the application of such MMs, asymmetric radiation can be successfully realized. An approximate closed-form expression for estimating the radiation asymmetry was derived by utilizing the MM's intrinsic impedance. It is evident that the occurrence of magnetic plasmons (or the condition of $\mu_x \approx 0$) is crucial for maximizing the radiation asymmetry of such a system in practice. It is also revealed that the radiation asymmetry is subject to internal bounds and may not exceed unity for a given bianisotropy.

We emphasize that the asymmetric system discussed here is rather different from the twisted system made of chiral metamaterials. Such as the one reported in Referece [49], wherein a TE polarized wave is converted to a TM polarized one (or vice versa), after travelling through the proposed metamaterial slab; while the transmission asymmetry is observed that both sides of the slab are illuminated by plane waves with identical polarization (both TE, or both TM). Noticing the twisting effect, it is in fact that there are two polarization systems involved. And, the transmission asymmetry will be reversed if the polarization of incidence is changed, e.g. from TE to TM, or from TM to TE. In particular, if the incidence is circularly polarized, asymmetric transmission cannot be observed. Our work presents the asymmetric radiation without any twist effect, i.e. only one polarization system is considered and discussed. Noticing the duality of TE and TM polarizations, our work would facilitate the realization of asymmetric radiation from arbitrary polarized sources, since the metamaterial can be easily designed to work for both the TE and TM cases.

Our analysis can be extended to other scenarios, such as thin films or metasurfaces consisting of asymmetric constituents, provided that a similar ME resonance occurs. The asymmetric radiation system exhibited here is different from those diffractive metasurfaces, such as the one exhibited in Reference [50], since the radiation asymmetry is mainly determined by the individual constituents of the metamaterial slab and less grating effect is involved, which would be much more flexible and feasible in practical designs and implementations. We emphasize that the MM depicted in Figure 5 is realizable in the high-frequency range. For instance, a facile multilayer electroplating fabrication technique may facilitate the preparation of metasurfaces made of standing-up SRRs in the terahertz range.[51] Another example in the optical range can be found in Reference [26], in which the forward-backward emission symmetry is broken by an array of pyramid-shaped nanoparticles with ME resonances. Hence, various potential applications can be expected. Our work reveals the physical mechanism of asymmetric radiation and provides a design principle for systems that exhibit such asymmetry, which is expected to facilitate the implementation of directional optical and quantum devices such as emitters and lasers as well as, by the principle of reciprocity, absorbers or solar cells optimized for unidirectional illumination.

**Supporting Information**

Supporting Information is available from the Wiley Online Library or from the author.

**Acknowledgements**

This work was sponsored by the National Natural Science Foundation of China under Grants No. 61372022, No. 61625502, No. 61574127, No. 61601408, No. 61775193 and No. 11704332, the ZJNSF under Grant No. LY13F010020 and No. LY17F010008, the Top-Notch Young Talents Program of China, the Fundamental Research Funds for the Central Universities under Grant No. 2017XZZX008-06, and the Innovation Joint Research Center for Cyber-Physical-Society System. B.Z. acknowledges funding support from Singapore Ministry of Education (MOE2015-T2-1-070, MOE2016-T3-1-006, Tier 1 RG174/16(S)) and Nanyang Technological University (NAP Start-Up Grant and Nanyang Research Award (Young Investigator)).

**Figures:**

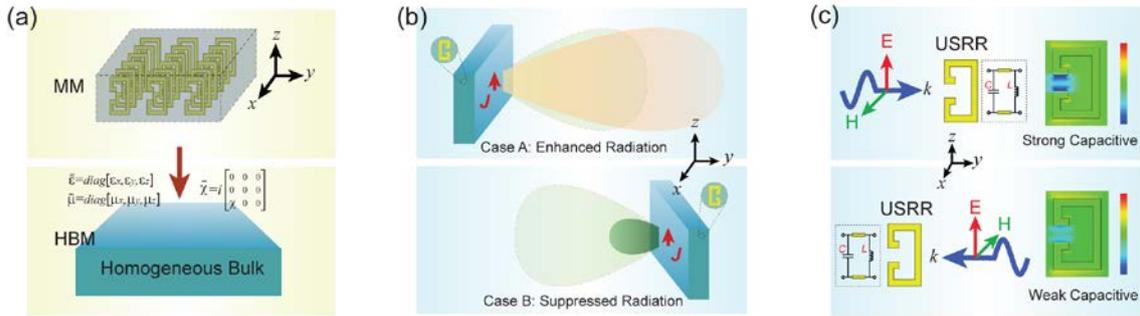

**Figure 1.** (a) A bianisotropic MM consisting of standard USRRs. (b) Schematic illustrations of the radiation asymmetry produced by a current in the presence of a bulk MM. The green shaded pattern with a red dashed outline represents the radiation of *J* alone. (c) The asymmetrical EM response of a USRR. The $E_z$ distributions are shown in the figure. For better interpretation, a simplified circuit model of the USRR is provided. Although this model with only passive elements cannot fully represent the magneto-electric resonator, it helps the understanding of the network effect that should not be neglected in such an asymmetric system. In these two simulations, the incident plane waves have identical amplitudes.

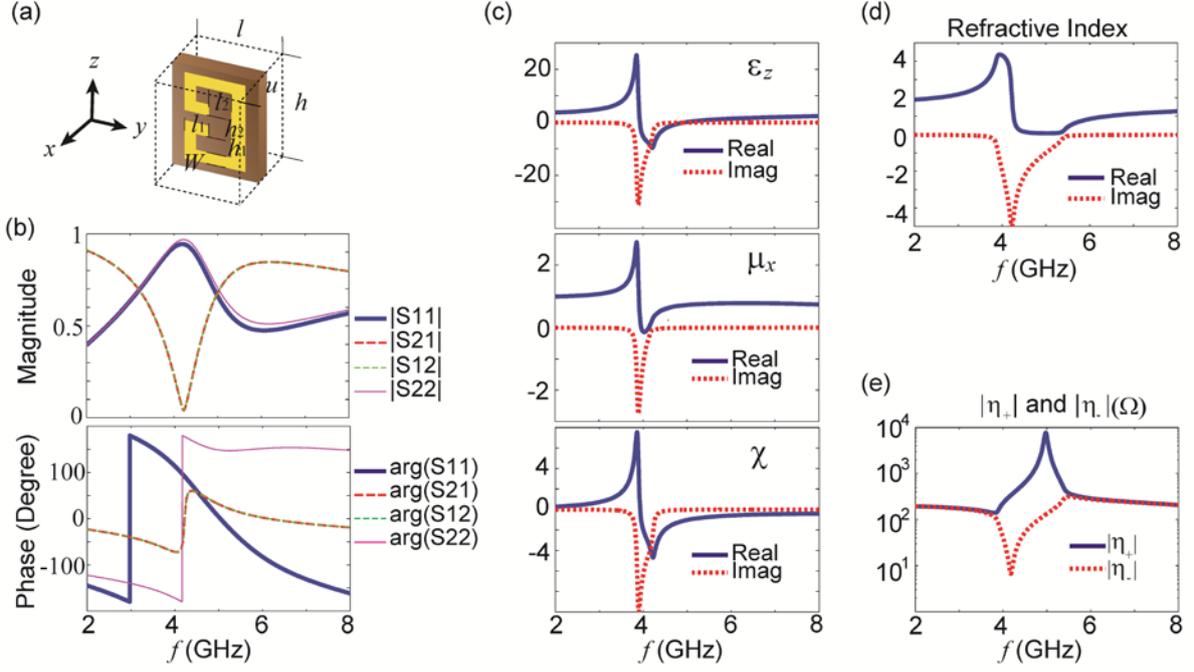

**Figure 2.** (a) The unit cell structure of the bianisotropic MM. In this design, $u=10$ mm, $l=8$ mm, $h=11$ mm, $l_1=2$ mm, $l_2=3$ mm, $h_1=1.5$ mm, $h_2=3$ mm, and $W=1$ mm. The USRR is made of copper with a thickness of 0.035 mm. (b) The simulated magnitudes and phases of the S-parameters for a MM slab consisting of a single-layer USRR array. (c) The extracted effective parameters ($\varepsilon_z$, $\mu_x$ and $\chi$). (d) The refractive index in the region of interest. The index is almost purely imaginary from below 4.2 GHz to above 5.4 GHz. (e) The magnitudes of the intrinsic wave impedances ($|\eta_+|$ and $|\eta_-|$) evaluated using the retrieved constitutive parameters.

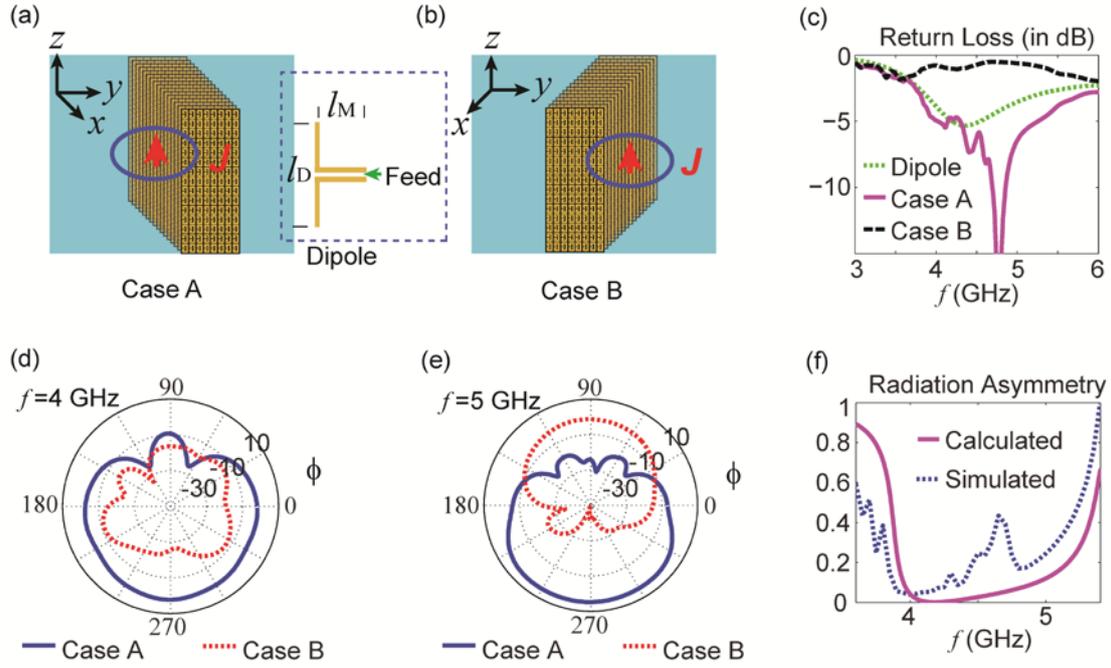

**Figure 3**. (a-b) Two cases illustrating control over the radiation from a short current. In the simulations presented here, the bulk MM is made of USRRs such as those shown in Figure 2. The unit cell periods in the *x*, *y* and *z* directions are 16, 8 and 15, respectively. In Case A/B, a short dipole with a section of matching transmission line (see the inset) is placed to the left/right center of the MM slab to excite a current. (c) The return losses of the dipole system alone and in Cases A and B. (d-e) The radiated power patterns in Cases A and B at 4 GHz and 5 GHz, respectively. (f) The radiation asymmetry in the band of interest. Solid line: calculated from Equation (1); dashed line: extracted from full-wave simulations. In deriving the radiation asymmetry, the gain of the antenna system was used to represent the radiated power level in each case.

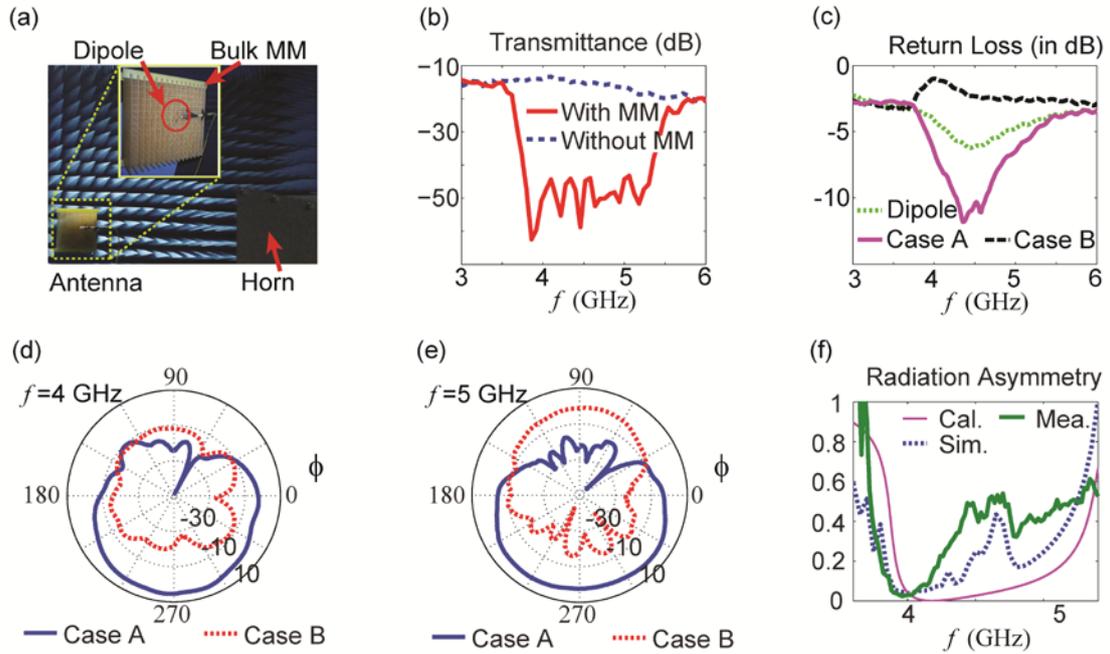

**Figure 4.** (a) The experimental setup for measuring the radiation from the MM-assisted antenna. The two insets show the fabricated dipole and the bulk MM. A standard horn antenna was used to receive the radiated power. (b) The results of transmittance measurements performed to test the forbidden band of the bulk MM. For these measurements, two horn antennas were placed facing each other in the chamber, and the bulk MM was placed in front of the receiving horn. (c) The return losses for the dipole antenna system alone and with the MM back reflector, i.e., in Cases A and B. The dipole is seen to be well matched in Case A but mismatched in Case B. (d-e) The measured radiation patterns in the two MM cases at 4 GHz and 5 GHz. The radiation in Case A is obviously much more efficient than that in Case B. (f) The measured radiation asymmetry. To facilitate comparison, the theoretical and simulated radiation asymmetry curves are also reproduced here.

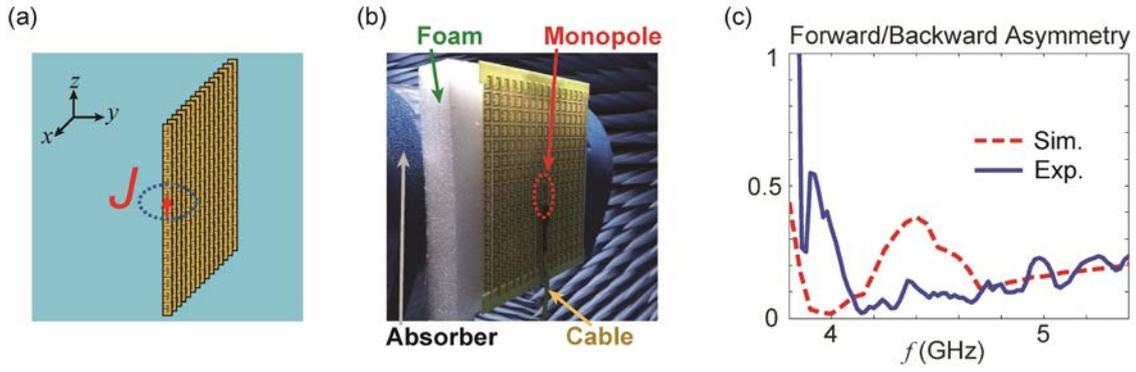

**Figure 5.** (a) Schematic design for the generation of asymmetric radiation by means of an ultra-thin MM consisting of a one-layer USRR array. The source current is placed at the center of the one-layer structure. (b) The experimental setup used to measure the radiation asymmetry of an antenna consisting of a monopole and an ultra-thin MM. The antenna was supported by an absorber together with a foam structure (the white structure in the photograph), with negligible influence on the results. For clarity, both the monopole and the feed cable are indicated. (c) The simulated and measured forward-to-backward ($+\hat{y}$ to $-\hat{y}$) radiation asymmetry. Although the frequency response is slightly shifted, the dip in the radiation asymmetry appears as expected.